# Zero Decomposition with Multiplicity of Zero-Dimensional Polynomial Systems

Yinglin Li, Bican Xia,* Zhihai Zhang

School of Mathematical Sciences

Peking University, Beijing 100871, China

**Abstract**

We present a zero decomposition theorem and an algorithm based on Wu's method, which computes a zero decomposition with multiplicity for a given zero-dimensional polynomial system. If the system satisfies some condition, the zero decomposition is of triangular form.

**Key words** polynomial system, zero decomposition, multiplicity, Wu's method

## 1 Introduction

Since Wu's method [8] was invented, many algorithms for zero decomposition of polynomial systems have been proposed [1, 4, 6, 7, 9]. The so-called *zero decomposition* is to decompose the zero set defined by a polynomial set into a union of finite many constructible sets. Usually, these constructible sets are defined by some *triangular sets*, i.e., the equations defining each constructible set are in *triangular form*, e.g., $f_1(x_1), f_2(x_1, x_2), \ldots, f_n(x_1, ..., x_n)$. The zeros of such equations can be described and solved easily, so the methods of zero decomposition have been applied successfully to many problems with various backgrounds, especially those in geometry.

All the existing methods for zero decomposition focus on the decomposition of the set of zeros and do not care about the multiplicities of (isolated) zeros. For a zero-dimensional polynomial system, we are interested in getting its zero decomposition with multiplicity. More precisely, for a zero-dimensional polynomial system $S$, we hope we can compute finite many pairs $[T_i, g_i]$ such that

$$\mathrm{MZero}(S) = \bigcup_i \mathrm{MZero}(T_i/g_i),$$

where $\mathrm{MZero}(\cdot)$ stands for the set of zeros (counted with multiplicity) and $\mathrm{MZero}(T_i/g_i) = \{\xi : \forall f \in T_i, f(\xi) = 0, g_i(\xi) \neq 0\}$ (counted with multiplicity). Herein, $T_i$ is a polynomial set, $g_i$ is a polynomial and for all $i \neq j$, $\mathrm{MZero}(T_i/g_i) \cap \mathrm{MZero}(T_j/g_j) = \varnothing$. That is to say, for any zero $p$ of $S$, there exists only one pair $[T_j, g_j]$ such that $p$ is a zero of $T_j$ and $g_j(p) \neq 0$ and the multiplicity of $p$ as a zero of $S$ is the same as that of it as a zero of $T_j$. We will call such decomposition *keep multiplicity*.

For a known zero $p$ of a zero-dimensional system $S$, one can compute the multiplicity of $p$ through Gröbner bases computation [2]. Furthermore, one can even compute the multiplicity structure of $p$ by the method in [3] using dual space. Obviously, our interest is different from those in [2] and [3]. We hope we can compute a zero decomposition which keeps multiplicity for a zero-dimensional system by modifying Wu's method. In that way, we will not add too much additional heavy computation to the original method. Our interest is also different from that in [5], where a new definition of multiplicity based on non-standard analysis was given and was proven to be equivalent to the classical definition. Zero decomposition is not a topic in [5].

---

*Corresponding author, email: xbc@math.pku.edu.cn

We hope that a zero decomposition keeping multiplicity is also triangular. If so, the multiplicities of zeros of each component can be read out directly [10]. In this paper, we give a zero decomposition theorem and an algorithm based on Wu's method, which computes a zero decomposition with multiplicity for a given zero-dimensional polynomial system. If the system satisfies some assumption, the zero decomposition is of triangular form.

The organization of the paper is as follows. In Section 2, some preliminary knowledge about the definition of multiplicity is reviewed. Section 3 proves a theorem of zero decomposition with multiplicity for given zero-dimensional polynomial systems. Section 4 presents the algorithm of zero decomposition with multiplicity and points out the condition for the algorithm to output triangular decomposition. We discuss some possible improvements on the algorithm in Section 5.

## 2 Preliminaries

First, let's recall the definition of *local (intersection) multiplicity*. We follow the notations in [2]. Although some notations and definitions can be stated in a more general way, we restrict ourselves to the ring $\mathbb{C}[X] = \mathbb{C}[x_1, \ldots, x_n]$ since we are interested in the complex or real zeros of zero-dimensional polynomial systems.

For $p = (\eta_1, \ldots, \eta_n) \in \mathbb{C}^n$, we denote by $M_p$ the maximal ideal generated by $\{x_1 - \eta_1, \ldots, x_n - \eta_n\}$ in $\mathbb{C}[X]$, and write

$$\mathbb{C}[X]_{M_p} = \left\{ \frac{f}{g} : f, g \in \mathbb{C}[X], g(\eta_1, \ldots, \eta_n) \neq 0 \right\}.$$

It is well-known that $\mathbb{C}[X]_{M_p}$ is the so-called *local ring*.

**Definition 1** *Let $I$ be a zero-dimensional ideal in $\mathbb{C}[X]$, and assume that $p \in \text{Zero}(I)$, the zero set of $I$ in $\mathbb{C}$. Then the* multiplicity *of $p$ as a point in $\text{Zero}(I)$ is defined to be*

$$\dim_{\mathbb{C}} \mathbb{C}[X]_{M_p} / I\mathbb{C}[X]_{M_p},$$

*the dimension of the quotient space $\mathbb{C}[X]_{M_p} / I\mathbb{C}[X]_{M_p}$ as a vector space in $\mathbb{C}$.*

There is another equivalent definition of local multiplicity using dual space.

For every index array $j = [j_1, \ldots, j_n] \in \mathbb{N}^n$, define the differential operator

$$\partial_j \equiv \partial_{j_1 \cdots j_n} \equiv \partial_{x_1^{j_1} \cdots x_n^{j_n}} \equiv \frac{1}{j_1! \cdots j_n!} \frac{\partial^{j_1 + \cdots + j_n}}{\partial x_1^{j_1} \cdots \partial x_n^{j_n}}$$

and the order of $\partial_j$ is $|\partial_j| = \sum_{l=1}^{n} j_l$.

Consider a system of $m$ polynomials, $\{f_1(X), \ldots, f_m(X)\}$, in $n$ variables with an isolated zero $\xi \in \mathbb{C}^n$, where $m \geq n$. Let $I = \langle f_1(X), \ldots, f_m(X) \rangle$. A functional is defined at $\xi \in \mathbb{C}^n$ as follows:

$$\partial_j[\xi] : \mathbb{C}[X] \to \mathbb{C},$$

where

$$\partial_j[\xi](f) = (\partial_j f)(\xi), \ \forall f \in \mathbb{C}[X].$$

All functionals at $\xi$ that vanish on $I$ form a vector space $D_\xi(I)$,

$$D_\xi(I) \equiv \{v = \sum_{j \in \mathbb{N}^n} v_j \partial_j[\xi] : \ v(f) = 0, \forall f \in I\}$$

where $v_j \in \mathbb{C}$. The vector space $D_\xi(I)$ is called the *dual space* of $I$ at $\xi$. For $\alpha = 0, 1, \ldots$, $D_\xi^\alpha(I)$ consists of functionals in $D_\xi(I)$ with order bounded by $\alpha$.

**Definition 2** *[3] The local multiplicity of zero $\xi$ of a zero-dimensional ideal $I \subseteq \mathbb{C}[X]$ is $m$ if the dual space $D_\xi(I)$ is of dimension $m$.*

**Remark 1** *According to [3], Definition 2 is equivalent to Definition 1.*

**Proposition 1** *[3] Let $\sigma$ be the smallest $\alpha$ such that $\dim(D_\xi^\alpha) = \dim(D_\xi^{\alpha+1}(I))$, then $D_\xi(I) = D_\xi^\sigma$. Furthermore we have*

$$\sigma < m = \dim(D_\xi(I)).$$

**Lemma 1** *Assume $\xi$ is an isolated zero of $I$ and the local multiplicity of $\xi$ is $m$. If $\xi \in \text{Zero}(g(X))$ where $g(X) \in \mathbb{C}[X]$, then for any $v \in D_\xi(I)$, any $h(X) \in \mathbb{C}[X]$ and $l \geq m$, we have $v(h(X)g(X)^l) = 0$.*

PROOF. If $\partial_j \equiv \partial_{j_1 \cdots j_n} \in D_\xi(I) = D_\xi^\sigma(I)$, $|\partial_j| \leq \sigma < m$ by Proposition 1. Because $g(X)^{l-|\partial_j|}$ is a factor of $\partial_j(hg^l)$, $\partial_j[\xi](h(X)g(X)^l) = 0$ since $\xi \in \text{Zero}(g(X))$. Because any $v \in D_\xi(I)$ is a linear combination of $\partial_j \in D_\xi(I) = D_\xi^\sigma(I)$, we are done.

## 3 Zero decomposition with multiplicity

Existing algorithms for triangular decomposition of zero-dimensional polynomial systems do not care about the multiplicities of the systems' zeros. For example, Wu's method does not keep the local multiplicities of zeros. We will prove in this section that for a given zero-dimensional polynomial system $S$, there is an algorithm which computes finite many pairs $[T_i, g_i]$ where $T_i$ is a characteristic set and $g_i$ is a polynomial such that

$$\text{MZero}(S) = \bigcup_i \text{MZero}(T_i/g_i)$$

as long as $S$ satisfies some special conditions. Herein, $\text{MZero}(\cdot)$ is the set of zeros counted with multiplicity. Also, we have

$$\text{MZero}(T_i/g_i) \cap \text{MZero}(T_j/g_j) = \emptyset \text{ if } i \neq j.$$

We mainly follow Wu's method and make necessary modifications for our purpose. Throughout this paper, we use without definition some basic concepts and notations from [8] with which we assume the readers are familiar.

For two polynomials $f(x)$ and $g(x)$, $\mathrm{prem}(f, g, x)$ is the *pseudo remainder* of $f$ pseudo-divided by $g$ with respect to (w.r.t.) $x$. For a polynomial $f(X)$ and a triangular set $T$, $\mathrm{prem}(f, T)$ is the *(successive) pseudo remainder* of $f$ w.r.t. $T$. If $r = \mathrm{prem}(f, T)$, it is clear that $r \in \langle f, T \rangle$, the ideal generated by $\{f\} \cup T$.

According to Wu's method, for a given polynomial system $S$, one can compute a so-called *characteristic set* of $S$, say $C = [C_1(X), \ldots, C_r(X)]$, such that

$$\mathrm{Zero}(S) = \mathrm{Zero}(C/J) \cup \bigcup_{i=1}^{r} \mathrm{Zero}(S, I_i), \tag{1}$$

where $I_i$ is the *initial (leading coefficient)* of $C_i(X)$ for $i = 1, \ldots, r$, $J = \prod_{j=1}^{r} I_j$, $\mathrm{Zero}(C/J) = \mathrm{Zero}(C) \setminus \mathrm{Zero}(J)$ and $\mathrm{Zero}(S, I_i) = \mathrm{Zero}(S \cup \{I_i\})$. Throughout this paper, we name the algorithm in [8] for computing the characteristic set $C$ in (1) as `WuCharSet`.

The following lemma shows that the local multiplicity of each point in $\mathrm{Zero}(C/J)$ is unchanged. That is to say, the first characteristic set computed by `WuCharSet` keeps multiplicity unchanged.

**Lemma 2** *Assume $S = \{f_1(X), \ldots, f_s(X)\}$, $I = \langle S \rangle$ is a zero-dimensional ideal, and $C = [C_1(X), \ldots, C_r(X)]$ is a characteristic set of $S$ computed by Wu's method such that (1) holds. Then for each $p \in \mathrm{Zero}(C/J)$, the local multiplicity of $p$ as a point in $\mathrm{Zero}(I)$ and the local multiplicity of $p$ as a point in $\mathrm{Zero}(\langle C \rangle)$ are the same.*

PROOF. Assume $p = (p_1, \ldots, p_n) \in \mathrm{Zero}(C/J)$, and let $M_p = \langle x_1 - p_1, \ldots, x_n - p_n \rangle$. According to Definition 1, the local multiplicity of $p$ as a point in $\mathrm{Zero}(I)$ is $\dim_{\mathbb{C}} \mathbb{C}[x_1, \ldots, x_n]_{M_p} / I\mathbb{C}[x_1, \ldots, x_n]_{M_p}$ and the local multiplicity of $p$ as a point in $\mathrm{Zero}(\langle C \rangle)$ is $\dim_{\mathbb{C}} \mathbb{C}[x_1, \ldots, x_n]_{M_p} / \langle C \rangle \mathbb{C}[x_1, \ldots, x_n]_{M_p}$. It suffices to prove

$$I\mathbb{C}[x_1, \ldots, x_n]_{M_p} = \langle C \rangle \mathbb{C}[x_1, \ldots, x_n]_{M_p}.$$

On one hand, $\langle C \rangle \mathbb{C}[x_1, \ldots, x_n]_{M_p} \subseteq I\mathbb{C}[x_1, \ldots, x_n]_{M_p}$ since $\langle C \rangle \subseteq I$. On the other hand, $\mathrm{prem}(f_i(X), C) = 0$ for each $f_i(X) \in S$ since $C$ is a characteristic set of $S$. Therefore, there exist $a_1, \ldots, a_r, q_1(X), \ldots, q_r(X)$ such that

$$\prod_{j=1}^{r} I_j^{a_j} f_i = \sum_{j=1}^{r} q_j(X) C_j(X).$$

Because $p \in \mathrm{Zero}(C/J)$, $I_j(p) \neq 0$ for all $1 \leq j \leq r$. Thus, each $I_j(X)$ is invertible in $\mathbb{C}[x_1, \ldots, x_n]_{M_p}$. As a result, $f_j(X) \in \langle C \rangle \mathbb{C}[x_1, \ldots, x_n]_{M_p}$. Thus,

$$I\mathbb{C}[x_1, \ldots, x_n]_{M_p} \subseteq \langle C \rangle \mathbb{C}[x_1, \ldots, x_n]_{M_p}.$$

That completes the proof.

For those points in $\mathrm{Zero}(S, I_i)$, their local multiplicities as points of $\mathrm{Zero}(S, I_i)$ may be not equal to their local multiplicities as points of $\mathrm{Zero}(S)$. Intuitively, if $p \in \mathrm{Zero}(S, I_i)$ and its local multiplicity as a point of $\mathrm{Zero}(S)$ is $m_1$, its local multiplicity as a point of $\mathrm{Zero}(S, I_i^{m_2})$ should also be $m_1$ provided that $m_2 \geq m_1$. And this is indeed true as implied by the following lemma.

**Lemma 3** *Suppose $\xi \in \text{Zero}(S, g(X))$ is an isolated zero of $S = \{f_1(X), \ldots, f_s(X)\}$ and $m$ is the local multiplicity of $\xi$ as a point of $\text{Zero}(S)$. Then the local multiplicity of $\xi$ as a point of $\text{Zero}(S, g(X)^l)$ is $m$ if $l \geq m$.*

PROOF. Let $I = \langle S \rangle$ and $I' = \langle S, g(X)^l \rangle$. According to Definition 2, it is sufficient to prove that
$$D_\xi(I) = D_\xi(I').$$

Obviously, $D_\xi(I) \supseteq D_\xi(I')$ according to the definition of dual space since $I \subseteq I'$.

On the other hand, for any $v \in D_\xi(I)$ and $f = \sum_{j=1}^{s} d_j(X) f_j(X) + h(X) g(X)^l \in I'$,
$$v(f(X)) = v(\sum_{j=1}^{s} d_j(X) f_j(X)) + v(h(X) g(X)^l) = v(h(X) g(X)^l)$$

because $\sum_{j=1}^{s} d_j(X) f_j(X) \in I$. According to Lemma 1, $v(h(X) g(X)^l) = 0$ and thus $v \in D_\xi(I')$. That completes the proof.

By Lemma 2, Lemma 3 and eq.(1), the following theorem is obvious.

**Theorem 1** *(Zero Decomposition with Multiplicity) Suppose $S$ is a finite non-empty set of non-zero polynomials in $\mathbb{C}[X]$ and $\langle S \rangle$ is a zero-dimensional ideal. Assume that $C = [C_1(X), \ldots, C_n(X)]$ is a characteristic set of $S$, $I_i$ is the initial of $C_i(X)$ and $J(i) = \prod_{k=1}^{i} I_k$ for $1 \leq i \leq n$. There exists $e_i \in \mathbb{N}$ for $2 \leq i \leq n$ such that*
$$\text{MZero}(S) = \text{MZero}(C/J(n)) \cup \bigcup_{i=2}^{n} \text{MZero}(S_i/J(i-1)), \qquad (2)$$

*where $S_i = S \cup \{I_i^{e_i}\}$, is a disjoint decomposition of $\text{MZero}(S)$.*

The existence of such $e_i$s is clear. For example, we can let $e_i$ be the Bézout bound, or set $e_i \geq m$ where $m = \dim_\mathbb{C} \mathbb{C}[x_1, \ldots, x_n]/\langle S \rangle$ is the number (counted with multiplicity) of points in $\text{MZero}(S)$, which can be obtained by computing the Gröbner basis of $\langle S \rangle$[2].

Additional, the characteristic set $C$ in (2) satisfies $\langle S \rangle = \langle S \cup C \rangle$, so we replace $S$ by $S \cup C$ sometimes to make sure our algorithm terminates.

## 4 Algorithm and examples

For a given $S$ in Theorem 1, the characteristic set in eq.(2) can be computed through `WuCharSet`. Naturally, one may hope to apply `WuCharSet` to each $S_i$ recursively to obtain a triangular decomposition of $\text{MZero}(S)$. However, if we directly apply `WuCharSet` to $S_i$ or $S_i \cup C$ recursively, the process may not terminate.

We take the following strategy. Let $S'_i = S \cup C \cup \{r_i\}$ where $r_i = \text{prem}(I_i^{e_i}, C)$. We replace $S_i$ by $S'_i$ in the recursive process. Because $r_i$ is reduced w.r.t. $C$, the *rank* of the characteristic set of $S'_i$ must be less than that of $C$ provided that $r_i \neq 0$.

However, we have to prove that this modification does not change the local multiplicity of each point in $\text{Zero}(S_i/J(i-1)) = \text{Zero}(S'_i/J(i-1))$.

**Proposition 2** *Let the notations be as above. Then for each $i \geq 2$ and a point $p \in \text{Zero}(S_i/J(i-1))$, the local multiplicity of $p$ as a point of $\text{Zero}(S_i)$ is equal to its local multiplicity as a point in $\text{Zero}(S_i')$.*

PROOF. Assume that $p = (p_1, \ldots, p_n)$ and $M_p = \langle x_1 - p_1, \ldots, x_n - p_n \rangle$. It is sufficient to prove that
$$\langle S_i \rangle \mathbb{C}[X]_{M_p} = \langle S_i' \rangle \mathbb{C}[X]_{M_p}.$$
Since $C$ is a characteristic set of $S$, $\langle C \rangle \subseteq \langle S \rangle$. $I_i$ is a polynomial in $x_1, \ldots, x_{i-1}$ because $\langle S \rangle$ is zero-dimensional. Then there exist $n_1, \ldots, n_{i-1} \in \mathbb{N}$ and $q_1(X), \ldots, q_{i-1}(X) \in \mathbb{C}[X]$ such that
$$\prod_{j=1}^{i-1} I_j^{n_j} I_i^{e_i} = \sum_{j=1}^{i-1} q_j(X) C_j(X) + r_i.$$
On one hand, it is obvious that $r_i \in \langle S_i \rangle$, so
$$\langle S_i \rangle \mathbb{C}[X]_{M_p} \supseteq \langle S_i' \rangle \mathbb{C}[X]_{M_p}.$$
On the other hand, $I_j (1 \leq j \leq i-1)$ is invertible in $\mathbb{C}[X]_{M_p}$ because $I_j(p) \neq 0$ when $j < i$. Then $I_i^{e_i} \in \langle S_i' \rangle \mathbb{C}[X]_{M_p}$. Thus
$$\langle S_i \rangle \mathbb{C}[X]_{M_p} \subseteq \langle S_i' \rangle \mathbb{C}[X]_{M_p}.$$
That completes the proof.

**Corollary 1** *Let notations be as above. For $2 \leq i \leq n$, $\text{MZero}(S_i/J(i-1)) = \text{MZero}(S_i'/J(i-1))$. Then we can rewrite eq.(2) as*
$$\text{MZero}(S) = \text{MZero}(C/J(n)) \cup \bigcup_{i=2}^{n} \text{MZero}(S_i'/J(i-1)). \tag{3}$$

Now, we may compute, for each $S_i'$, a characteristic set and obtain a zero decomposition with multiplicity of $\text{MZero}(S_i')$. Finally, by the discussion before Proposition 2, the process must terminate as long as $r_i = \text{prem}(I_i^{e_i}, C) = 0$ never occurs. Then we have a triangular decomposition with multiplicity of the original polynomial system.

The main algorithm can be outlined below.

**Algorithm: ZeroDecompMulti**
Input: polynomial system $PS$ which satisfies $\langle PS \rangle$ is a zero-dimensional ideal.
Output: a zero decomposition with multiplicity of $\text{MZero}(PS)$.

1. Initialize SET1=$\{[PS, 1]\}$, SET2=$\varnothing$ and SET3=$\varnothing$.

2. Estimate a bound on the number of zeros (counted with multiplicity) of $PS$, denoted by $m$.

3. Repeat the following steps until SET1=$\varnothing$

   3.1 Choose $[S, P]$ in SET1 and set SET1 = SET1 $\setminus \{[S, P]\}$.

   3.2 For the polynomial system $S$, compute its characteristic set by Wu's method. If it's inconsistent, go to Step 3.1. Otherwise, denote the characteristic set by $C = [C_1(X), \ldots, C_n(X)]$, and let $I_i$ be the leading coefficient of $C_i(X)$, $J(i) = \prod_{k=1}^{i} I_k$ $(1 \leq i \leq n)$.

3.3 Let SET2 = SET2 ∪ {[C, P × J(n)]}.

3.4 For all $i$ ($2 \leq i \leq n$), compute $r_i = \text{prem}(I_i^m, C)$. If $r_i = 0$, let SET3 = SET3 ∪ {[S ∪ C, P × J(i − 1)]} and SET2 = SET2 \ {[C, P × J(n)]}, else let SET1 = SET1 ∪ {[S ∪ C ∪ $r_i$, P × J(i − 1)]}.

4. Output SET2 and SET3.

If $r_i \neq 0$ always occur, the rank of the characteristic set of new polynomial system must be less than that of the original one. So the termination of the algorithm is obvious. Otherwise, we put the new polynomial system into SET3 as a part of the output. Eventually, we get a zero decomposition with multiplicity of $PS$:

$$\text{MZero}(PS) = \bigcup_{[T,P] \in SET2} \text{MZero}(T/P) \cup \bigcup_{[S,Q] \in SET3} \text{MZero}(S/Q) \qquad (4)$$

Now we present an example to demonstrate the algorithm.

**Example 1** *[3] Consider a system of 3 polynomials $PS = \{x^2 + y + z − 1, x + y^2 + z − 1, x + y + z^2 − 1\}$. We try to compute a zero decomposition with multiplicity of $PS$. First, we use the Bézout bound on the number of zeros, $m = 8$, which is the product of all total degree of polynomials in $PS$. Next, we compute a characteristic set of $PS$:*

$$C = [x^2(x^2 + 2x − 1)(x − 1)^2, x^2(x^2 + 2y − 1), x^2(x^2 − 1 + 2z)].$$

*According to Theorem 1, $\text{MZero}(PS) = \text{MZero}(C/x^4) \cup \text{MZero}(PS1)$, where $PS1 = PS \cup C \cup \{x^{16}\}$. We continue to compute a zero decomposition of $\text{MZero}(PS1)$. The characteristic set of $PS1$ is:*

$$C1 = [x^2, y − x − y^2, 1 − z − y].$$

*Therefore $\text{MZero}(PS1) = \text{MZero}(C1)$. Finally, we get a zero decomposition with multiplicity of $\text{MZero}(PS)$ as follows:*

$$\text{MZero}(PS) = \text{MZero}(C/x^4) \cup \text{MZero}(C1).$$

*Because $r_i = 0$ never happens during the process, the decomposition is triangular.*

Now, we know that $PS$ can be decomposed with multiplicity into 3 branches:

$$\begin{aligned}T_1 &= [x^2 + 2x − 1, x^2 + 2y − 1, x^2 − 1 + 2z],\\ T_2 &= [(x − 1)^2, x^2 + 2y − 1, x^2 − 1 + 2z],\\ T_3 &= [x^2, y − x − y^2, 1 − z − y].\end{aligned}$$

The multiplicity of each zero in $T_1$ is 1, and the zeros in $\text{Zero}(T_2)$ or $\text{Zero}(T_3)$ all have multiplicity 2.

The following example points out that some polynomial systems cannot be decomposed into triangular form through the algorithm `ZeroDecompMulti`.

**Example 2** *[3] Consider a system of three polynomials $PS = \{x^3 − yz, y^3 − xz, z^3 − xy\}$. The Bézout bound on the number of the zeros is $m = 27$. We compute a characteristic set of $PS$:*

$$C = [x^7 − x^{11}, yx^4 − yx^8, −zx^5 + zx^9].$$

*According to Theorem 1, we know* $\mathrm{MZero}(PS) = \mathrm{MZero}(C/(x^4 - x^8)(x^9 - x^5)) \cup \mathrm{MZero}(PS \cup C \cup (x^4 - x^8)^{27}) = \mathrm{MZero}(C/(x^4 - x^8)(x^9 - x^5)) \cup \mathrm{MZero}(PS \cup C)$ *since* $\mathrm{prem}((x^4 - x^8)^{27}, C) = 0$. *For* $PS1 = PS \cup C$, *its characteristic set is also* $C$, *so algorithm* `ZeroDecompMulti` *cannot get a further decomposition of* $\mathrm{MZero}(PS1)$. *Hence, the algorithm gives us a decomposition of* $\mathrm{MZero}(PS)$ *as follows:*

$$\mathrm{MZero}(PS) = \mathrm{MZero}(C/(x^4 - x^8)) \cup \mathrm{MZero}(PS \cup C).$$

Actually, there is no zeros in $\mathrm{MZero}(C/(x^4 - x^8))$, so the decomposition is meaningless.

## 5 Improvements on the algorithm

We discuss some possible improvements on the algorithm `ZeroDecompMulti` in this section.

First, we consider the bound on the number of zeros of the original polynomial system, which is set to be a constant $m$ in the algorithm `ZeroDecompMulti`. Obviously, the smaller the $m$ is, the easier the computation will be.

Intuitively, we can set $m$ to be the number of zeros (counted with multiplicity) of the original polynomial system, which can be obtained by computing the Gröbner basis of the system. But we will have some additional time cost for computing the Gröbner basis, which violates our intention obviously. So obtaining a reasonable bound according to the form or characteristic of the polynomial system with little cost is more intelligent.

Another strategy is to define $m$ as a variable, and update it during the decomposition process. Review eq.(2), suppose that the number of zeros of polynomial system $S$ does not exceed $m_0$, then that of $\bigcup_{i=2}^{n} \mathrm{MZero}(S_i/J(i-1))$ must be no more than $m_0 - |\mathrm{MZero}(C/J(n))|$. Here $|\mathrm{MZero}(C/J(n))|$ means the number of zeros which satisfy $C = 0$ and $J(n) \neq 0$ (counted with multiplicity). As a result, we could make the following adjustment in the algorithm `ZeroDecompMulti`: when adding $[PS, P]$ into SET2, set $m = m - |\mathrm{MZero}(PS/P)|$. Usually, $|\mathrm{MZero}(PS/P)|$ is easy to calculate, so the computation will be more efficient.

Second, we consider the situation that $r_i = \mathrm{prem}(I_i^m, C) = 0$ occurs. Since the characteristic set of $S_i'$ computed by `WuCharSet` must be $C$, we could not get a further decomposition of $\mathrm{MZero}(S_i'/J(i-1))$ by the algorithm `ZeroDecompMulti` and only record this situation in SET3. We may try the following strategy.

Assume that $r_i = \mathrm{prem}(I_i^m, C) = 0$ and $\overline{C_i(X)} = C_i(X) - I_i \times x_i^{\mathrm{ldeg}(C_i(X))} \neq 0$, then we replace $S_i$ in eq.(2) by $S_i^\sharp = S \cup C \cup \{\overline{C_i(X)}^m\}$. The proposition below proves that the multiplicities of zeros are still unchanged.

**Proposition 3** *Let the notations be as above. If* $r_i = \mathrm{prem}(I_i^m, C) = 0$, *then*

$$\mathrm{MZero}(S_i/J(i-1)) = \mathrm{MZero}(S_i^\sharp/J(i-1)).$$

PROOF. Since $r_i = \mathrm{prem}(I_i^m, C) = 0$, it's easy to prove that

$$\mathrm{MZero}(S_i/J(i-1)) = \mathrm{MZero}(S \cup C \cup \{I_i^m\}/J(i-1)) = \mathrm{MZero}(S \cup C/J(i-1)).$$

Because $\mathrm{Zero}(S \cup C/J(i-1)) = \mathrm{Zero}(S_i^\sharp/J(i-1))$, we know from Lemma 3 that

$$\mathrm{MZero}(S \cup C/J(i-1)) = \mathrm{MZero}(S_i^\sharp/J(i-1)).$$

That completes the proof.

Now we can try to decompose $S_i^\sharp$ as long as $\overline{C_i(X)} \neq 0$. Let

$$r_i' = \text{prem}(\overline{C_i(X)}^m, C),$$

it's easy to prove that the conclusion still hold if we replace $\overline{C_i(X)}^m$ in Proposition 3 by $r_i'$. Then, the rank of the characteristic set of $S_i^\sharp$ must be less than that of $C$, as long as $r_i' \neq 0$.

**Example 3** *Consider a system of three polynomials*

$$S = \{x^2 + y, 4xy + 2y^2, (x+y)z^2 + z + 1\}.$$

*According to the algorithm* `ZeroDecompMulti`, *we compute a characteristic set of $S$: $C = [-2x^3 + x^4, x^2 + y, (x - x^2)z^2 + z + 1]$, and get a decomposition*

$$\text{MZero}(S) = \text{MZero}(C/x - x^2) \cup \text{MZero}(S_1),$$

*where $S_1 = S \cup C \cup \{(x-x^2)^{12}\}$. Since $\text{prem}((x-x^2)^{12}, C) = 512x^3 \neq 0$, we compute a characteristic set of $S_1' = S \cup C \cup \{x^3\}$: $C_1 = [x^3, x^2+y, (x-x^2)z^2+z+1]$. So the decomposition of $S_1$ is*

$$\text{MZero}(S_1) = \text{MZero}(C_1/x - x^2) \cup \text{MZero}(S_2),$$

*where $S_2 = S_1' \cup C_1 \cup \{(x-x^2)^{12}\}$. Clearly, $\text{MZero}(C_1/x - x^2) = \varnothing$. The decomposition process will terminate because $\text{prem}((x-x^2)^{12}, C_1) = 0$. However, this is not a triangular decomposition. Now, we use the method based on Proposition 3, replacing $S_2$ by $S_2^\sharp = S_1' \cup C_1 \cup \{(z+1)^{12}\}$ before computing the characteristic set. Finally we get a characteristic set $C_2 = [x^3, x^2+y, (-1+22x-232x^2)z - 1 + 21x - 211x^2]$ of $S_2^\sharp$ and a complete triangular decomposition of $S$:*

$$\text{MZero}(S) = \text{MZero}(C/x - x^2) \cup \text{MZero}(C_2).$$

Finally, we take into account the case of redundant components in the zero decomposition. Obviously, the decomposition will have a redundant component if some leading coefficients of the characteristic set have the same prime factors. Therefore, we should factorize these leading coefficients and consider the factors one by one. The multiplicities of zeros remain unchanged as long as the condition in Lemma 3 holds. This strategy may let us get a better decomposition sometimes.

**Example 4** *Review the polynomial system in Example 2:*

$$PS = \{x^3 - yz, y^3 - xz, z^3 - xy\}.$$

*A characteristic set of $PS$ is $C = [x^7 - x^{11}, yx^4 - yx^8, -zx^5 + zx^9]$, the corresponding leading coefficients are $I = [-1, x^4 - x^8, -x^5 + x^9]$. All the square-free factors in $I$ are $x, 1-x, 1+x, 1+x^2$. So we know that $PS$ has a zero decomposition with multiplicity as below:*

$$\begin{aligned}
\text{MZero}(PS) = \ & \text{MZero}(PS \cup \{x^{27}\}) \cup \\
& \text{MZero}(PS \cup \{(1-x)^{27}\}/x) \cup \\
& \text{MZero}(PS \cup \{(1+x)^{27}\}/x(1-x)) \cup \\
& \text{MZero}(PS \cup \{(1+x^2)^{27}\}/x(1-x^2)).
\end{aligned}$$

*After decomposing each part of it, we have the final result:*

$$\begin{aligned}
\text{MZero}(PS) = \quad & \text{MZero}(PS \cup \{x^7, x^4y, x^5z\}) \cup \\
& \text{MZero}([1-x, -y^4+1, -y^3+z]) \cup \\
& \text{MZero}([x+1, y^4-1, -y^3-z]) \cup \\
& \text{MZero}([1+x^2, y^4-1, y^3-xz]).
\end{aligned}$$

*Although there is still one component which is not triangularized, we find 3 triangular sets that keep the multiplicities unchanged. This is a better result compared to that of Example 2.*

# Acknowledgement

This work is supported in part by NSFC-90718041 and NKBRPC-2005CB321902 in China.